\documentclass[%
 reprint,
 amsmath,amssymb,
 aps,
longbibliography,
onecolumn,
preprint
]{revtex4-1}

\usepackage[dvipdfmx]{graphicx}
\usepackage{dcolumn}
\usepackage{bm}
\usepackage{natbib}
\usepackage{color}

\begin{document}


\title{Energy transport due to pressure diffusion enhanced by helicity and system rotation in inhomogeneous turbulence}

\author{Kazuhiro Inagaki}
 \email{kinagaki@iis.u-tokyo.ac.jp}
\author{Fujihiro Hamba}%
\affiliation{%
 Institute of Industrial Science, The University of Tokyo, Tokyo, Japan
}%

%


\date{\today}




\allowdisplaybreaks[1]

\begin{abstract}

It is known that turbulent energy is rapidly transferred in the direction of the rotation axis in a rotating system, in comparison with the non-rotating case. In this study, this phenomenon is investigated as a problem of energy diffusion expressed by the Reynolds averaged Navier-Stokes (RANS) model. The conventional gradient-diffusion approximation for the turbulent energy flux cannot account for the enhanced energy transport observed in rotating inhomogeneous turbulence. In experiments, inhomogeneity of turbulence is modeled with an oscillating grid, leading to the turbulent energy falling off away from the grid. In order to adequately describe the phenomenon, we propose a new model for the energy flux due to the pressure associated with the rotational motion of a fluid. The model of the energy flux is expressed to be proportional to the turbulent helicity, which is the statistically averaged value of the inner product of the velocity and vorticity fluctuations. This property is closely related to the group velocity of inertial waves in a rapidly rotating fluid. The validity of the model is assessed using a direct numerical simulation (DNS) of inhomogeneous turbulence under rotation where the flow configuration is similar to the oscillating-gird turbulence. It is shown that most of the turbulent energy transport enhanced by the system rotation is attributed to the pressure diffusion term. The spatial distribution of the energy flux due to the pressure related to the system rotation is similar to that of the turbulent helicity with negative coefficient. Hence, the new model which is proportional to the turbulent helicity is able to qualitatively account for the enhanced energy flux due to the system rotation. Finally, the helical Rossby number is proposed in order to estimate the relative importance of the energy flux enhanced by the turbulent helicity and the rotation, in comparison to the conventional gradient-diffusion approximation.

\end{abstract}

\pacs{Valid PACS appear here}
\maketitle


\section{\label{sec:level1}Introduction}

Turbulent flows are known to be significantly affected by rotation. Many geophysicists, astrophysicists, meteorologists, and engineers are interested in the effects of rotation on turbulence. In a rapidly rotating fluid, the well known Taylor-Proudman theorem suggests that a flow tends to become two-dimensional. Although an exactly two-dimensional flow is not always established especially in wall-confined flows, the tendency of the quasi two-dimensionalization is observed in homogeneous turbulence under rotation, using the eddy-damped quasi-normal Markovian (EDQNM) approximation \cite{cj1989}, the large-eddy simulation (LES) \cite{cambonetal1997}, and direct numerical simulation (DNS) \cite{mnr2001,ymk2011}. These results indicate that the inter-scale energy transfer in the wavenumber space is altered due to the system rotation. In the case of decaying homogeneous turbulence, system rotation prevails against the energy cascade to the small scale, and the decay rate of the turbulent energy is reduced \cite{bfr1985,mnr2001}. Since the energy cascade is the inter-scale energy transfer, it may be difficult to be modeled in one-point closure. In order to avoid this difficulty, the reduction of the dissipation rate of the turbulent energy is modeled in the Reynolds averaged Navier-Stokes (RANS) modeling instead of modeling the reduction of the cascade rate. As a result, a term accompanied with the rotation is added in the transport equation for the turbulent energy dissipation rate \cite{bfr1985,okamoto1995}. Although this is an indirect modeling of the phenomenon, the reduction of decay rate of the turbulent energy can be predicted by the one-point closure model of the RANS equation.

The RANS models are more often applied to inhomogeneous turbulence. In the previous studies of the RANS modeling, the effects of the system rotation on inhomogeneous turbulence are mainly discussed in terms of the Reynolds stress. In these studies, the effects of the system rotation on the Reynolds stress are expressed in the form of the nonlinear eddy-viscosity models with rotation-dependent coefficients \cite{syb2000,wj2000} and with the turbulent helicity \cite{yy1993,yb2016,inagakietal2017}. Here, the turbulent helicity is defined as $H = \langle u_i' \omega_i' \rangle$ where $\langle \rangle$ denotes the ensemble average, and $u_i'$ and $\omega_i'$ are the velocity and vorticity fluctuations, respectively. Note that the turbulent helicity is not the total amount in volume but the statistically averaged value at one point, so that it can vary in time and space. On the other hand, the effects of the system rotation on the turbulent energy transport were not discussed. This might be because the Coriolis force does not perform work on the fluid, and the turbulent energy transport equation is not altered.

\begin{figure}[htp]
\centering
\includegraphics[scale=0.32]{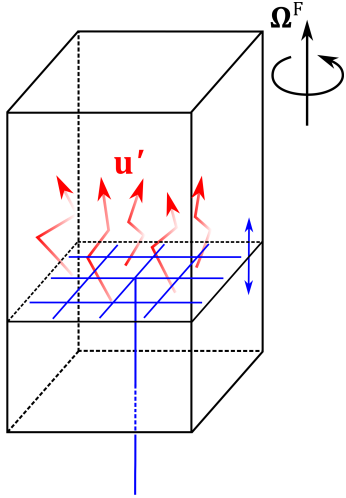}
\caption{Schematic diagram for oscillating-grid turbulence. $\mathbf{\Omega}^\mathrm{F}$ denotes the angular velocity of the system rotation.}
\label{fig:1}
\end{figure}

In order to focus on the effects of rotation on the turbulent energy transport rather than on the Reynolds stress, it is useful to consider the inhomogeneous flow fields with zero mean velocity. A simple example of such a flow is oscillating-grid turbulence \cite{ht1976,dl1978}. Its schematic flow configuration is shown in Fig.~\ref{fig:1}, in which the system rotation rate $\mathbf{\Omega}^\mathrm{F}$ is zero. In the experiment involving this flow, the turbulent energy is generated by an oscillating grid in a tank and spatially transferred in one direction perpendicular to the grid plane. Dickinson and Long \cite{dl1978} experimentally suggested that the diffusion of the turbulent energy can be represented by the `eddy viscosity,' and the width of the turbulence region around the grid $d$ grows as $d \sim t^{1/2}$. This suggestion indicates that the diffusion in oscillating-grid turbulence can be predicted using the conventional gradient-diffusion approximation. Matsunaga \textit{et al}. \cite{matsunagaetal1999} revealed that the spatial distribution of the turbulent energy in a steady state of the oscillating-grid turbulence can be predicted by the conventional $K$-$\varepsilon$ model.

These results suggest that the gradient-diffusion approximation with the eddy viscosity is suitable for the description of inhomogeneous turbulence without rotation. However, this is not the case for rotating turbulence. Dickinson and Long \cite{dl1983} performed an experiment involving oscillating-grid turbulence with system rotation where the axis was perpendicular to the grid plane (Fig.~\ref{fig:1}). They revealed that the width of the turbulence region grows as $d \sim t$; the growth is faster than the non-rotating case where $d \sim t^{1/2}$. The same result was obtained by experiments \cite{dsd2006,kolvinetal2009} and a numerical simulation \cite{rd2014}. This fact suggests that for rotating inhomogeneous turbulence, the diffusion of the turbulent energy cannot be simply described by the gradient-diffusion approximation since the time dependence of the width of the turbulence region is notably different from the well-known diffusion problem. In other words, the phenomenon cannot be predicted by the conventional RANS models using the gradient-diffusion approximation with the eddy viscosity. Moreover, Godefered and Lollini \cite{gl1999} performed the DNS which mimics the rotating oscillating-grid turbulence and showed that the spatial distribution of the turbulent energy is significantly affected by system rotation. Although Yoshizawa \cite{yoshizawa2002} proposed a model for the pressure-velocity correlation associated with the mean rotational motion of a fluid, this model represents the energy flux in the direction perpendicular to the rotation axis. Hence, this model cannot account for the energy transfer enhanced in the direction parallel to the rotation axis.

Ranjan and Davidson \cite{rd2014} discussed the relationship between the growth of the width of the turbulence region and the inertial wave observed in rotating oscillating-grid turbulence in detail. The group velocity of the inertial wave is mostly directed to the rotation axis and its sign corresponds to the sign of the instantaneous helicity, $u_i' \omega_i'$ \cite{moffatt1970}; the wave packets with negative (positive) instantaneous helicity propagate in the positive (negative) direction of the rotation axis. In fact, it was showen the instantaneous helicity is successively segregated in a statistical sense in their simulation \cite{rd2014}, which is consistent with the inertial wave propagation. These facts imply that in the RANS modeling, the energy transport enhanced in rotating oscillating-grid turbulence can be described in terms of the turbulent helicity. Inagaki \textit{et al}. \cite{inagakietal2017} showed that in the case of inhomogeneous helical turbulence subject to system rotation, the pressure diffusion term significantly contributes to the Reynolds stress transport. They also confirmed that the correlation between the velocity and the pressure fluctuation can be expressed by the product of the turbulent helicity and the absolute vorticity vector. In contrast to the model proposed by Yoshizawa \cite{yoshizawa2002}, the model proposed by Inagaki \textit{et al}. \cite{inagakietal2017} represents the energy flux in the direction parallel to the rotation axis. Thus, this model is expected to account for the enhanced energy transport in the direction parallel to the rotation axis for rotating inhomogeneous turbulence.

Helical flow structures are often seen in engineering fields such as a swirling flow in a straight pipe \cite{kitoh1991,steenbergen} or the swirling jet \cite{stepanovetal2018}, and meteorological flows including supercell \cite{lilly1986,nn2010}. It is possible that the turbulent helicity affects the energy flux in such flows. The model suggested by Inagaki \textit{et al}. \cite{inagakietal2017} may be useful for predicting such helical flows with rotation. In the context of magnetohydrodynamics (MHD), the turbulent helicity is known to be essential for the alpha dynamo effect \cite{moffattbook,krauseradler,bs2005}. The model of the energy flux accompanied with the turbulent helicity and rotation suggests that the turbulent helicity affects not only the mean magnetic field, but also the turbulent kinetic energy. The prediction of the turbulent kinetic energy is significant for the estimation of the turbulent time scale in the alpha coefficient in the RANS models of the MHD turbulence \cite{hamba2004,wby2016,wby2016-2}.

In this study, we assess the validity of the model proposed by Inagaki \textit{et al}. \cite{inagakietal2017} which is associated with the turbulent helicity by using a DNS of freely decaying inhomogeneous turbulence with and without system rotation whereby the rotation axis is parallel to the inhomogeneous direction of turbulence. The simulation configuration is similar to that proposed by Ranjan and Davidson \cite{rd2014} in which the simulation of rotating turbulence is performed starting from a spatially confined homogeneous isotropic turbulence. We focus on the RANS modeling of the phenomenon in low to moderate rotation cases. This is because under such circumstances, neither the conventional gradient-diffusion approximation nor the linear inviscid solution of the Navier-Stokes equation is suitable for the description of the flow. Low to moderate rotation cases correspond to fully developed turbulent flows, and they are the target of the RANS modeling. In this study, the transport equation for the turbulent energy is examined. The validity of the newly proposed model is discussed compared to the simulation result. Finally, the helical Rossby number is proposed as a criterion for judging the relative importance of the enhanced energy flux due to the turbulent helicity and the rotation in general turbulent flows.

The organization of this paper is as follows. In Sec~\ref{sec:level2}, the turbulent energy transport and the conventional model approach are described. In addition, a new model expression is proposed for the energy flux enhanced by the turbulent helicity and the rotation. In Sec.~\ref{sec:level3}, the numerical setup and the simulation results are presented. A discussion of the validity of the new model is presented in Sec.~\ref{sec:level4}. The helical Rossby number is proposed as a criterion for judging the relative importance of the energy flux enhanced by the turbulent helicity and rotation in general turbulent flows. Finally, a summary is provided and conclusions are discussed in Sec~\ref{sec:level5}.

\section{\label{sec:level2}Turbulent energy transport enhanced by the turbulent helicity and rotation}

The Navier-Stokes equation and the continuity equation for an incompressible fluid in a rotating system are given, respectively, by
\begin{align}
\frac{\partial u_i}{\partial t} & =
- \frac{\partial}{\partial x_j} u_i u_j - \frac{\partial p}{\partial x_i}
+ \nu \nabla^2 u_i
+ 2 \epsilon_{ij\ell} u_j \Omega^\mathrm{F}_\ell
+ f^\mathrm{ex}_i,
\label{eq:1} \\
\frac{\partial u_i}{\partial x_i} & = 0, 
\label{eq:2}
\end{align}
where $u_i$ is the $i$th component of velocity, $p$ the pressure divided by the fluid density with the centrifugal force included, $\nu$ the kinematic viscosity, $\nabla^2 (=\partial^2/\partial x_j \partial x_j)$ the Laplacian operator, $\Omega^\mathrm{F}_i$ the angular velocity of the system rotation, $\epsilon_{ij\ell}$ the alternating tensor, and $f^\mathrm{ex}_i$ the external forcing. The pressure is determined by the following Poisson equation:
\begin{align}
\nabla^2 p = - s_{ij} s_{ij} + \frac{1}{2} \omega_i \omega_i + 2 \omega_i \Omega^\mathrm{F}_i,
\label{eq:3}
\end{align}
where $s_{ij} [ = (\partial u_i/\partial x_j + \partial u_j/\partial x_i )/2]$ is the strain rate of velocity and $\omega_i ( = \epsilon_{ij\ell} \partial u_\ell/\partial x_j )$ is the vorticity. In a non-rotating frame, the first two terms on the right-hand side of Eq.~(\ref{eq:3}) remain. The third term on the right-hand side of Eq.~(\ref{eq:3}) denotes the effect of the system rotation on the pressure. In order to directly evaluate the effects of rotation on the pressure, we decompose the pressure into a nonlinear component and a rotational component as $p = p^\mathrm{N} + p^\Omega$ where they are respectively defined as
\begin{subequations}
\begin{align}
\nabla^2 p^\mathrm{N} & = - s_{ij} s_{ij} + \frac{1}{2} \omega_i \omega_i, 
\label{eq:4a} \\
\nabla^2 p^\Omega & = 2 \omega_i \Omega^\mathrm{F}_i.
\label{eq:4b} 
\end{align}
\end{subequations}
Hereafter, we refer to $p^\mathrm{N}$ as the nonlinear pressure and refer to $p^\Omega$ as the rotational pressure.

\subsection{\label{sec:level2a}Turbulent energy transport equation}

In this study, the energy transport phenomenon observed in both non-rotating and rotating oscillating-grid turbulence are discussed in terms of the RANS equation. We decompose the physical quantities $q [= (u_i, p, \omega_i)]$ into the mean and the fluctuation parts as
\begin{align}
q = Q + q', \ \ & Q = \left< q \right>,
\label{eq:5}
\end{align}
where $\left< \right>$ denotes the Reynolds or ensemble averaging. Substituting Eq.~(\ref{eq:5}) into Eqs.~(\ref{eq:1}) and (\ref{eq:2}), we can derive the equations for the mean velocity and the velocity fluctuation. Then, the equation for the turbulent energy $K (= \langle u_i' u_i' \rangle/2)$ is written as
\begin{align}
\frac{\partial K}{\partial t} + \frac{\partial}{\partial x_i} U_i K
= P^K - \varepsilon + T^K + \Pi^K + D^K + F^K.
\label{eq:6}
\end{align}
Here, $P^K$ is the production rate, $\varepsilon$ the dissipation rate, $T^K$ the turbulent diffusion, $\Pi^K$ the pressure diffusion, $D^K$ the viscous diffusion, and $F^K$ the work done by the external forcing. They are respectively defined as
\begin{subequations}
\begin{align}
P^K & = - R_{ij} S_{ij}, 
\label{eq:7a} \\
\varepsilon & = \nu \left< \frac{\partial u_i'}{\partial x_j} \frac{\partial u_i'}{\partial x_j} \right>,
\label{eq:7b} \\
T^K & = -\frac{\partial}{\partial x_i} \left< u_i' \frac{1}{2} u_j' u_j' \right>,
\label{eq:7c} \\
\Pi^K & = -\frac{\partial}{\partial x_i} \left< u_i' p' \right>,
\label{eq:7d} \\
D^K & = \nu \nabla^2 K,
\label{eq:7e} \\
F^K & = \left< u_i' f^\mathrm{ex}_i{}' \right>,
\label{eq:7f}
\end{align}
\end{subequations}
where $S_{ij} [ = (\partial U_i/\partial x_j + \partial U_j/\partial x_i )/2]$ denotes the strain rate of the mean velocity and $R_{ij} (= \langle u_i' u_j' \rangle)$ is the Reynolds stress. It should be noted that the angular velocity of the system rotation does not appear explicitly in Eqs.~(\ref{eq:6}) and (\ref{eq:7a})--(\ref{eq:7f}) since the Coriolis force does not perform work. However, the effect of the system rotation on the turbulent energy transport should be incorporated thorough the rotational pressure [Eq.~(\ref{eq:4b})]. Then, we decompose the pressure diffusion term as
\begin{align}
\Pi^K = \Pi^\mathrm{N} + \Pi^\Omega, 
\label{eq:8} 
\end{align}
where $\Pi^\mathrm{N}$ and $\Pi^\Omega$ are respectively defined as
\begin{subequations}
\begin{align}
\Pi^\mathrm{N} & = -\frac{\partial}{\partial x_i} \left< u_i' p^\mathrm{N}{}' \right>,
\label{eq:9a} \\
\Pi^\Omega & = -\frac{\partial}{\partial x_i} \left< u_i' p^\Omega{}' \right>.
\label{eq:9b} 
\end{align}
\end{subequations}
Hereafter, we refer to $\Pi^\mathrm{N}$ as the nonlinear pressure diffusion and refer to $\Pi^\Omega$ as the rotational pressure diffusion.

\subsection{\label{sec:level2b}Diffusion problem described in terms of the RANS equation}

In the case of oscillating-grid turbulence, there is no mean velocity and the flow is homogeneous in two directions and inhomogeneous in one direction. Hereafter, we take the direction of the flow inhomogeneity and the rotation axis to be $z$. Then, the equation for $K$ is written as
\begin{align}
\frac{\partial K}{\partial t} =
- \varepsilon - \frac{\partial}{\partial z} \left( \left< u_z' \frac{1}{2} u_i' u_i' \right> + \left< u_z' p^\mathrm{N}{}' \right>
+ \left< u_z' p^\Omega{}' \right> \right)
+ \nu \frac{\partial^2 K}{\partial z^2} + F^K.
\label{eq:10}
\end{align}
In this case, $F^K$ represents the energy injection due to the grid oscillation. Here, all terms except for the viscous diffusion term are unknown variables and need to be modeled. In the conventional RANS modeling, $\varepsilon$ is usually obtained by solving its transport equation, and the diffusion terms are modeled by the gradient-diffusion approximation as
\begin{align}
\left< u_i' \frac{1}{2} u_j' u_j' \right> + \left< u_i' p^\mathrm{N}{}' \right> + \left< u_i' p^\Omega{}' \right>
= - \frac{\nu^\mathrm{T}}{\sigma_K} \frac{\partial K}{\partial x_i},
\label{eq:11}
\end{align}
where $\nu^\mathrm{T}$ is the eddy-viscosity coefficient expressed by $\nu^\mathrm{T} = C_\nu K^2/\varepsilon$ in which $C_\nu$ and $\sigma_K$ are model constants. For high-Reynolds-number turbulence, the diffusion by the kinematic viscosity is negligible and the model is given by
\begin{align}
\frac{\partial K}{\partial t} =
- \varepsilon + \frac{\partial}{\partial z} \left( \frac{C_\nu}{\sigma_K} \frac{K^2}{\varepsilon} \frac{\partial K}{\partial z} \right) + F^K.
\label{eq:12}
\end{align}
This model accurately predicts a non-rotating oscillating-grid turbulence since the eddy viscosity represents the energy diffusion  due to turbulent mixing. The gradient-diffusion approximation for the energy flux is consistent with the experimentally observed growth of the width of the turbulence region $d$, where $d \sim t^{1/2}$ \cite{dl1978}. Moreover, Matsunaga \textit{et al}. \cite{matsunagaetal1999} revealed that the spatial distribution of the turbulent energy in a steady state of the oscillating-grid turbulence can be predicted using the RANS model described by the gradient-diffusion approximation. However, the model given by Eq.~(\ref{eq:11}) does not contain the effects of system rotation. As such, this model cannot account for the enhancement of the energy transport observed in rotating oscillating-grid turbulence \cite{dl1983,dsd2006,rd2014,kolvinetal2009,gl1999}. There are some elaborate RANS models in which the effects of system rotation are incorporated. For example, the effect of the reduction of the energy cascade is considered by modifying the transport equation for $\varepsilon$ \cite{bfr1985,okamoto1995}, and the rotation-dependent model coefficient is proposed with the aid of the algebraic Reynolds stress model procedure \cite{syb2000,wj2000}. Although these effects are essential for describing some effects of rotation on turbulence, they are insufficient to predict the energy transport enhanced in rotating oscillating-grid turbulence. This is because these models are based on the gradient-diffusion approximation; so that, they cannot account for the rapid growth of the width of the turbulence region as $d \sim t$, which was confirmed by several previous works \cite{dl1983,dsd2006,kolvinetal2009,rd2014}. Yoshizawa \cite{tsdia} developed a statistical closure theory for inhomogeneous turbulence which is called the two-scale direct-interaction approximation (TSDIA). Yoshizawa \cite{yoshizawa2002} proposed a model for the pressure diffusion term containing the effects of the mean shear and the mean rotation with the aid of the TSDIA. The model expression is written as
\begin{align}
\left< u_i' p' \right> 
= C_{KPS} \frac{K^3}{\varepsilon^2} S_{ij} \frac{\partial K}{\partial x_j}
+ C_{KP\Omega} \frac{K^3}{\varepsilon^2} W_{ij} \frac{\partial K}{\partial x_j},
\label{eq:13}
\end{align}
where $W_{ij} [ = (\partial U_i/\partial x_j - \partial U_j/\partial x_i )/2 - \epsilon_{ij\ell} \Omega^\mathrm{F}_\ell]$ is the mean absolute vorticity tensor, and $C_{KPS}$ and $C_{KP\Omega}$ are model constants. In the case where there is no mean velocity and the system is rotating, the model is given by
\begin{align}
\left< u_i' p' \right> 
= - C_{KP\Omega} \frac{K^3}{\varepsilon^2} \epsilon_{ij \ell} \Omega^\mathrm{F}_\ell \frac{\partial K}{\partial x_j}.
\label{eq:14}
\end{align}
In this model, the relationship $\langle u_i' p' \rangle \Omega^\mathrm{F}_i = 0$ holds; that is, the velocity-pressure correlation is orthogonal to the angular velocity vector of the rotation. The model given by Eq.~(\ref{eq:14}) represents the energy flux in the direction perpendicular to the rotation axis. Hence, it cannot account for the energy transport enhanced in the direction parallel to the rotation axis. In summary, previous models cannot account for the enhancement of the energy transport in the direction parallel to the rotation axis.

\subsection{\label{sec:level2c}A model for energy flux enhanced by the turbulent helicity and system rotation}

Inagaki \textit{et al}. \cite{inagakietal2017} showed that the pressure diffusion term significantly contributes to the Reynolds stress transport in rotating inhomogeneous turbulence accompanied with the turbulent helicity. In their work, the effect of the system rotation on the velocity-pressure fluctuation correlation was analytically obtained with the aid of the TSDIA \cite{tsdia}. Detailed calculations are provided in Appendix~\ref{sec:a}. As a result, we obtain the following model:
\begin{align}
\left< u_i' p^\Omega{}' \right> = - C_\Omega \frac{K^3}{\varepsilon^2} H 2 \Omega^\mathrm{F}_i,
\label{eq:15} 
\end{align}
where $H (= \langle u_i' \omega_i' \rangle)$ is the turbulent helicity and $C_\Omega$ is a model constant. Since the turbulent helicity is not the total amount in volume but the statistically averaged value at one point, it can vary in time and space. Therefore, the pressure diffusion due to the rotational pressure is expressed as
\begin{align}
\Pi^\Omega = \frac{\partial}{\partial x_i} \left( C_\Omega \frac{K^3}{\varepsilon^2} H 2 \Omega^\mathrm{F}_i \right).
\label{eq:16} 
\end{align}
As seen from Eqs.~(\ref{eq:6}) and (\ref{eq:7d}), the correlation between the velocity and the pressure fluctuation is interpreted as the energy flux due to the pressure. Hereafter, we refer to $\langle u_i' p^\Omega{}' \rangle$ as the rotational pressure flux. Equation~(\ref{eq:15}) indicates that the negative turbulent helicity, $H<0$, invokes an energy flux parallel to the rotation axis, while the positive turbulent helicity, $H>0$, invokes a flux anti-parallel to the rotation axis. This property corresponds to the group velocity of inertial waves; the wave packets with negative instantaneous helicity, $u_i' \omega_i' < 0$, propagate upward, while the packets with positive instantaneous helicity, $u_i' \omega_i' > 0$, propagate downward. Thus, this model of the rotational pressure flux plays a similar role to that of group velocity of inertial waves. The detailed expression of the group velocity of an inertial wave is given in Appendix~\ref{sec:b}. This model is expected to account for the enhancement of the energy transport observed in a rotating oscillating-grid turbulence \cite{dl1983,dsd2006,rd2014,kolvinetal2009,gl1999}. In fact, in the simulation of Ranjan and Davidson \cite{rd2014}, negative helicity is dominant in the upper side of the turbulent cloud, while positive helicity is dominant in the lower side, so that energy is transferred outward from the cloud. Moreover, the rotational pressure flux can be interpreted as the energy flux due to the inertial waves in linear inviscid limit. This point is discussed in Appendix~\ref{sec:c}.

It should be noted that the effects of rotation should appear not only in the case of solid body rotation of the system, but also in the case in which the non-trivial mean vorticity exists. From the viewpoint of the covariance of model expression \cite{ariki2015}, Eq.~(\ref{eq:15}) should be written as
\begin{align}
\left< u_i' p^\Omega{}' \right> = - C_\Omega \frac{K^3}{\varepsilon^2} H \Omega^\mathrm{A}_i,
\label{eq:17} 
\end{align}
where $\Omega^\mathrm{A}_i (= \Omega_i + 2\Omega^\mathrm{F}_i)$ denotes the mean absolute vorticity vector and $\Omega_i = \epsilon_{ij\ell} \partial U_\ell /\partial x_j$. Hence, this effect of the turbulent helicity on the energy flux must be important in predicting the turbulent energy distribution in helical flows such as a swirling flow in a straight pipe \cite{kitoh1991,steenbergen}, the swirling jet \cite{stepanovetal2018}, supercell \cite{lilly1986,nn2010}, and also in the RANS model of the MHD turbulence \cite{hamba2004,wby2016,wby2016-2}.

\section{\label{sec:level3}Numerical simulation}

\subsection{\label{sec:level3a}Numerical setup}

In order to assess the validity of the model given by Eq.~(\ref{eq:15}), a DNS of inhomogeneous turbulence subject to system rotation is performed. The flow configuration is similar to that proposed by Ranjan and Davidson \cite{rd2014}. The computational domain is $L_x \times L_y \times L_z = 2\pi \times 2 \pi \times 2\pi$ and the number of grid points is $512^3$. The pseudo-spectral method is used and the aliasing error is eliminated by using the phase-shift method. For time integration, the 3rd-order Runge-Kutta scheme is adopted for nonlinear term, while the viscous and the Coriolis terms are solved exactly using the integral factor technique \cite{integraltechnique}. The initial velocity field is given by a homogeneous isotropic turbulence confined around the $z=0$ plane, and the rotation axis is directed to the $z$-axis. The parameters for the simulations are shown in Table \ref{tb:1}. Here, the Reynolds number $\mathrm{Re}$ and the Rossby number $\mathrm{Ro}$ are respectively defined as
\begin{align}
\mathrm{Re} = \frac{K_0^2}{\nu \varepsilon_0}, \ \ 
\mathrm{Ro} = \frac{\varepsilon_0/K_0}{2 \Omega^\mathrm{F}} ,
\label{eq:18}
\end{align}
where $K_0 = K |_{z=0,t=0} (= 0.704)$, $\varepsilon_0 = \varepsilon |_{z=0,t=0} (= 1.24)$, and $\Omega^\mathrm{F}$ denotes the absolute value of the angular velocity of the system rotation.

\begin{table}[t]
\centering
\caption{Simulation parameters. In all runs, the kinematic viscosity is set to $\nu = 10^{-3}$ and the Reynolds number is $\mathrm{Re} = 400$.}
\begin{ruledtabular}
\begin{tabular}{ccc}
Run & $\Omega^\mathrm{F}$ & $\mathrm{Ro}$ \\ \hline
0 &  $0$ & $\infty$ \\
08 & $0.8$ & $1.10$ \\
1 &  $1$ & $0.880$ \\
2 &  $2$ & $0.440$ \\
5 &  $5$ & $0.176$ \\
\end{tabular}
\end{ruledtabular}
\label{tb:1}
\end{table}

In order to generate the solenoidal initial velocity, we perform a pre-computation of decaying homogeneous isotropic turbulence. For the initial condition of the pre-computation, the energy spectrum is set to $E(k) \propto k^4 \exp [ -2(k/k^\mathrm{p})^2]$ where $k^\mathrm{p} = 6$ and $K = \int_0^\infty \mathrm{d}k \ E(k) = 2$. Here, we use the velocity field $u^\mathrm{hit}_i$ at the time $\sqrt{2K/3}|_{t=0} k^\mathrm{p} t = 7.97$ at which the energy has been transferred to high-wavenumber region and the energy dissipation rate starts decaying. The stream function $\psi_i$ is introduced which satisfies $u^\mathrm{hit}_i = \epsilon_{ij\ell} \partial \psi_\ell / \partial x_j$ and $\nabla^2 \psi_i = 0$. Then, the initial velocity field of the main computation of the inhomogeneous turbulence $u^\mathrm{ini}_i$ is given by
\begin{align}
u^\mathrm{ini}_i = \epsilon_{ij\ell} \frac{\partial}{\partial x_j} \left[ g(z) \psi_\ell \right].
\label{eq:19} 
\end{align}
Here, $g(z)$ is a weighting function which confines the velocity field around $z=0$ and is set to $g(z) = \exp [ - (z/\sigma)^4]$ where $\sigma = L_z /8 = 0.785$. The integral length scale obtained from $u^\mathrm{hit}_i$ field is $L^\mathrm{int} [= 3\pi/4K \int_0^\infty \mathrm{d} k \ k^{-1} E(k)] = 0.516$. Therefore, the width of the confined turbulent region, $2\sigma = 1.57$, is three times as wide as the integral length scale $L^\mathrm{int}$. 

\subsection{\label{sec:level3b}Results}

In the following results, statistical quantities are obtained by averaging over the $x$-$y$ plane. Thus, the statistical quantities only depend on $z$ and $t$.

\subsubsection{\label{sec:level3b1}Spatial distribution of the turbulent energy}

\begin{figure}[htp]
\centering
\includegraphics[scale=0.65]{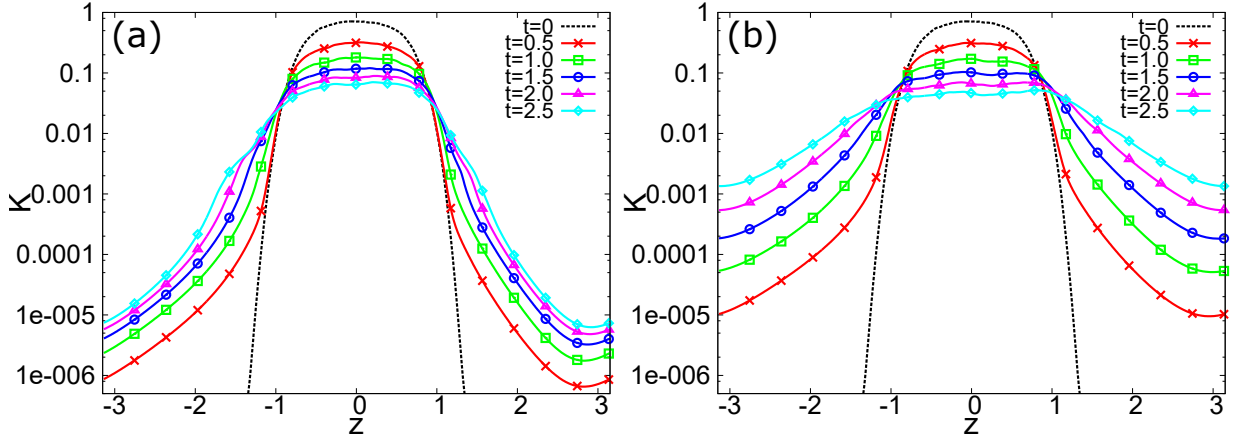}
\caption{Spatial distribution of turbulent energy at each time for (a) run 0 and (b) run 1.}
\label{fig:2}
\end{figure}

\begin{figure}[htp]
\centering
\includegraphics[scale=0.65]{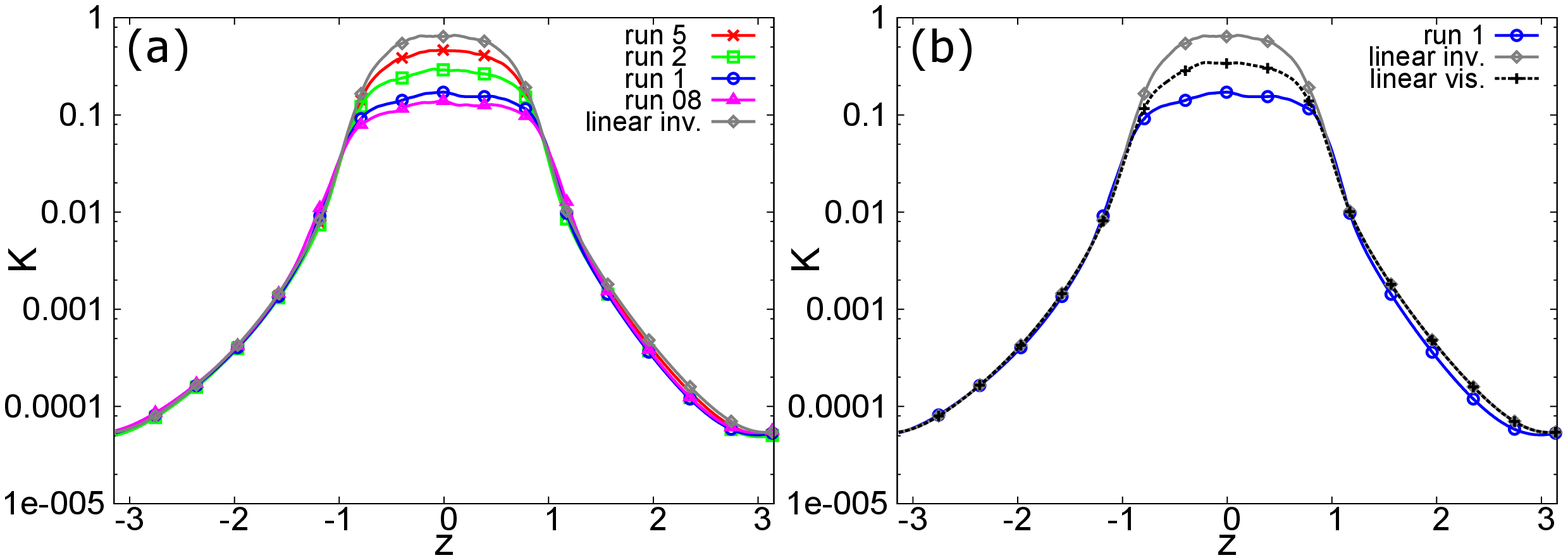}
\caption{Spatial distribution of the turbulent energy at $2\Omega^\mathrm{F}t = 2$: (a) comparison between four runs of the rotating cases with the linear inviscid solution (linear inv.) and (b) comparison between the numerical solution of run 1, the linear inviscid solution, and the solution of run 1 obtained from the Navier-Stokes equation without the nonlinear term (linear vis.).}
\label{fig:3}
\end{figure}

Figure \ref{fig:2} shows the spatial distribution of the turbulent energy at each time for runs 0 and 1. In both cases, the energy at $|z| < 1$ decreases while that at $|z| > 1$ increases with the progression of time. This represents the outward energy transfer. In the rotating case (run 1), the energy increase at $|z| > 1$ is faster and the resulting energy transfer is rapid compared with the non-rotating case (run 0). The spatial distribution of the turbulent energy at $2\Omega^\mathrm{F}t = 2$ for four runs of the rotating cases is shown in Fig.~\ref{fig:3}(a). Here, `linear inv.' denotes the solution of the linear inviscid equation given by Eq.~(\ref{eq:b1}). In the outer region at $|z| > 1$, all lines almost overlap, while in the center region at $|z| < 1$, the values are quite different from each other. It is seen that the spatial distribution of $K$ asymptotically approaches the linear inviscid solution as the rotation rate increases. In Fig.~\ref{fig:3}(b), we compare the numerical solution of the Navier-Stokes equation, the linear inviscid solution, and the linear viscous solution obtained from the Navier-Stokes equation without the nonlinear term for run 1, whose Rossby number is moderate. It is clearly seen that the linear viscous solution does not overlap with the numerical solution of the nonlinear Navier-Stokes equation. This result indicates that the nonlinearity is not negligible in the center region at $|z| < 1$ for the moderate-Rossby-number case.

\begin{figure}[htp]
\centering
\includegraphics[scale=0.63]{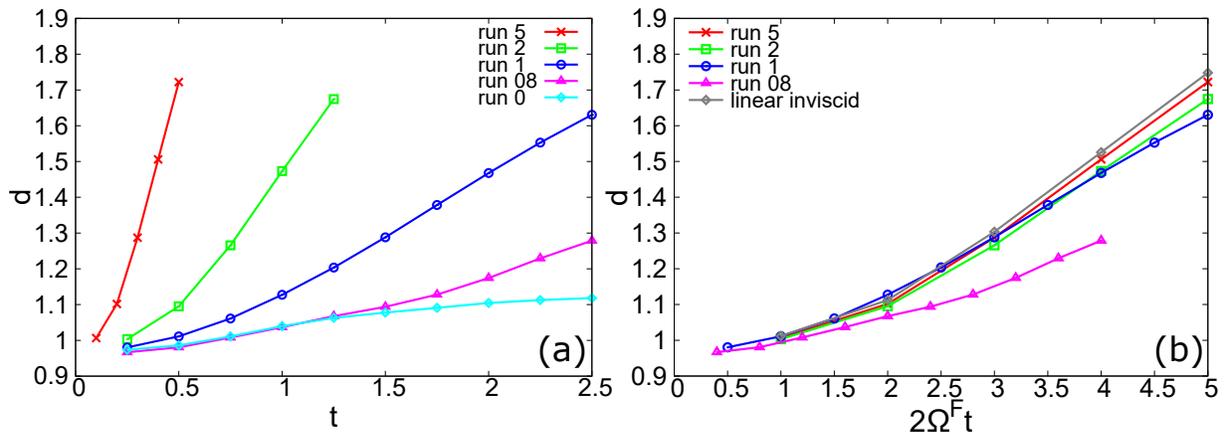}
\caption{The growth of the the width of the turbulence region against time. In (b), time is normalized by the angular velocity of the system rotation. The result of the linear inviscid case is also plotted.}
\label{fig:4}
\end{figure}

Dickinson and Long \cite{dl1983} suggested that the growth of the the width of the turbulence region $d$ varies according to the rotation rate. Here, we define the width of the turbulence region as $d = |z (K = 0.02 K_0)|$; that is, the location of the turbulence edge where the turbulent energy takes the value of $K = 0.02K_0$, which is similar to the previous method \cite{dl1983,dsd2006,kolvinetal2009,rd2014}. The time evolution of $d$ with time is shown in Fig.~\ref{fig:4}(a). The width of the turbulence region for the rotating cases appears to grow linearly, while that for the non-rotating case (run 0) is saturated at $t = 1.5$. The growth for run 0 is not exactly the same as the experimental result of $d \sim t^{1/2}$ \cite{dl1978} because of the absence of energy injection in this simulation. For the rotating cases, the growth rate increases as the rotation rate increases. The flux of the turbulent energy in the direction parallel to the rotation axis depends on the rotation rate. The previous studies showed that for rotating cases, the lines of growth of the width of the turbulence region overlap when time is normalized by the angular velocity of the system rotation \cite{dl1983,dsd2006,kolvinetal2009,rd2014}. Figure \ref{fig:4}(b) shows the growth of the width of the turbulence region against time normalized by the angular velocity of the system rotation. The linear inviscid solution is also plotted. In all of the rotating cases except for run 08, the lines overlap and are similar to the linear inviscid solution. This result suggests that the growth of the width of the turbulence region of rapidly rotating flows can be estimated by the linear inviscid equation given by Eq.~(\ref{eq:b1}). On the other hand, the gradient of the growth for run 08 is not as steep as those in the other runs in Fig.~\ref{fig:4}(b). This result indicates that not only the energy transport due to the rotation, but also the diffusion due to the nonlinearity of the turbulence is important for moderate-Rossby-number flows. 

\subsubsection{\label{sec:level3b2}Budget of the turbulent energy transport equation}

The budget of the turbulent energy transport equation (\ref{eq:10}) for runs 0, 1, and 5 at $2\Omega^\mathrm{F}t = 2$ ($t=1$ for run 0) is shown in Fig.~\ref{fig:5}. Note that $F^K = 0$, that is, the work done by the external forcing is zero in the simulation. In Fig.~\ref{fig:5}(a) for run 0, the turbulent energy dissipation is dominant at $|z|<1$ and the energy is transferred mainly by the turbulent diffusion given by Eq.~(\ref{eq:7c}) near $z = \pm 1$. On the other hand, in Fig.~{\ref{fig:5}(b) for run 1, the rotational pressure diffusion also contributes to the energy transfer at $z = \pm 1$. In Fig.~\ref{fig:5}(c) for run 5, the intensity of the rotational pressure diffusion is much larger than that of run 1 at the same time, where the time is normalized by the angular velocity of the system rotation. The turbulent energy increases at $1 < |z| < 2$ solely by the rotational pressure diffusion. As the Rossby number decreases, the contribution of the rotational pressure diffusion increases.

\begin{figure}[htp]
\centering
\includegraphics[scale=0.65]{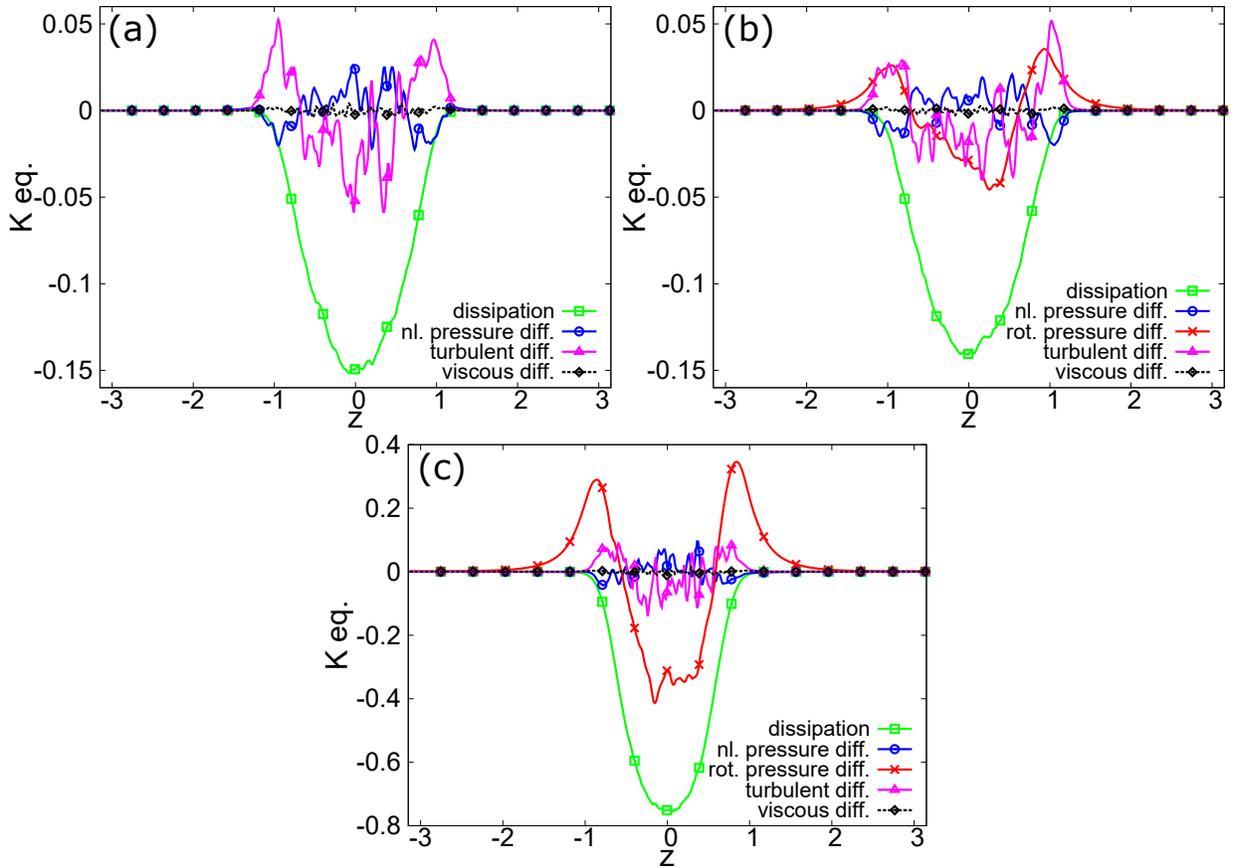}
\caption{The budget of the turbulent energy transport equation for (a) run 0 at $t=1$, (b) run 1 at $t=1 (2\Omega^\mathrm{F}t=2)$, and (c) run 5 at $t = 0.2 (2\Omega^\mathrm{F}t=2)$.}
\label{fig:5}
\end{figure}

\section{\label{sec:level4}Discussion}

\subsection{\label{sec:level4a}Evaluation of the model for the rotational pressure flux}

As seen in Fig.~\ref{fig:5}, the rotational pressure diffusion term significantly contributes to the budget of the turbulent energy transport. This result indicates that the energy flux due to turbulence is enhanced by the system rotation. Firstly, we examine whether the enhancement of the energy flux can be predicted by the conventional gradient-diffusion approximation. Figure~\ref{fig:6} shows the comparison between the total energy flux due to turbulence, $\langle u_z' p' \rangle + \langle u_z' u_i' u_i'/2 \rangle$, and the conventional gradient-diffusion approximation given by Eq.~(\ref{eq:11}) for runs 0 and 1 at $t=1$ and $2$ ($2\Omega^\mathrm{F}t=2$ and $4$ for run 1). The model constant is chosen as $C_\nu/\sigma_K = 0.22$ so that the agreement is good for run 0. The value of the model constant is almost twice as large as the conventional value $C_\nu/\sigma_K = 0.09$ \cite{matsunagaetal1999,yoshizawabook}. For run 0, the spatial distribution of the total energy flux can be predicted by the gradient-diffusion approximation although the model constant is large. On the other hand, for run 1, the energy flux is under-predicted by the gradient-diffusion approximation. In particular, a broad spatial distribution of the energy flux at $-3 < z < 3$ at $2\Omega^\mathrm{F} t = 4$ is not predicted. Figure~\ref{fig:7} shows the comparison between the energy flux due to the nonlinearity, $\langle u_z' p^\mathrm{N}{}' \rangle + \langle u_z' u_i' u_i'/2 \rangle$, and the gradient-diffusion approximation for runs 08 and 1 at $2\Omega^\mathrm{F}t=2$ and $4$. The model constant is the same as in Fig.~\ref{fig:6}. In contrast to Fig.~\ref{fig:6}(b), the energy flux due to the nonlinearity agrees fairly well with the gradient-diffusion approximation for both runs. Therefore, it is clearly shown that the conventional gradient-diffusion approximation can predict the energy flux due to the nonlinearity, but cannot account for the rotational pressure flux, $\langle u_z' p^\Omega{}' \rangle$, enhanced by the system rotation. Therefore, a new model is required to account for the rotational pressure flux.

\begin{figure}[htp]
\centering
\includegraphics[scale=0.65]{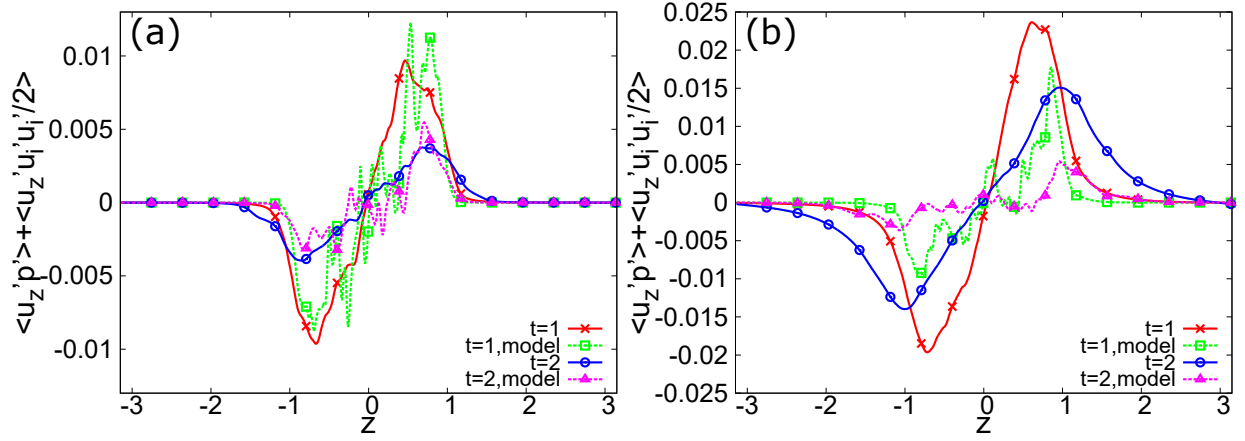}
\caption{Comparison between the total energy flux due to turbulence, $\langle u_z' p' \rangle + \langle u_z' u_i' u_i'/2 \rangle$, and the gradient-diffusion approximation given by Eq.~(\ref{eq:11}) for (a) run 0 and (b) run 1 at $t=1$ and $2$ ($2\Omega^\mathrm{F}t=2$ and $4$ for run 1).}
\label{fig:6}
\end{figure}

\begin{figure}[htp]
\centering
\includegraphics[scale=0.65]{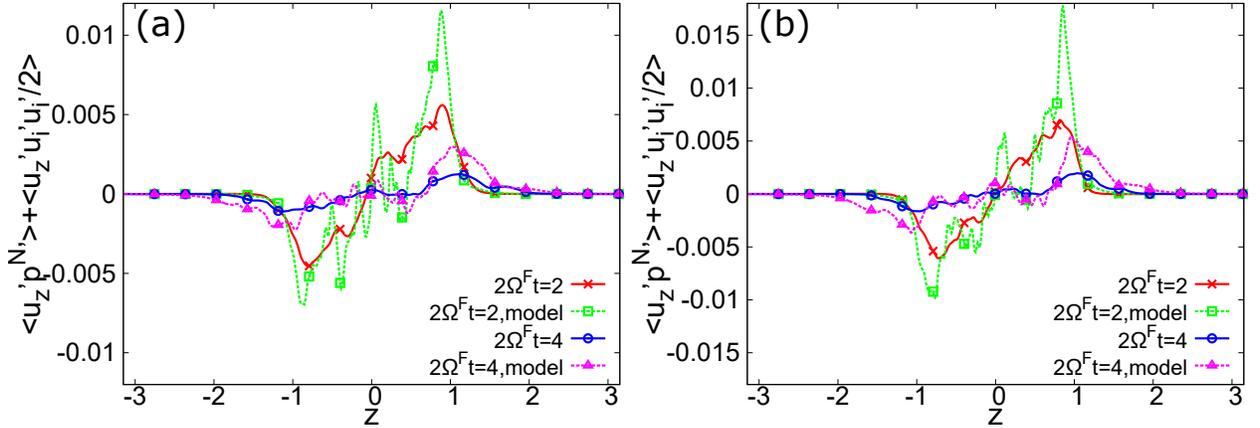}
\caption{Comparison between the energy flux due to the nonlinearity, $\langle u_z' p^\mathrm{N}{}' \rangle + \langle u_z' u_i' u_i'/2 \rangle$, and the gradient-diffusion approximation given by the right-hand side of Eq.~(\ref{eq:11}) for (a) run 08 and (b) run 1 at $2\Omega^\mathrm{F} t = 2$ and $4$.}
\label{fig:7}
\end{figure}

Next, we examine the newly proposed model given by Eq.~(\ref{eq:15}). Figures~\ref{fig:8}(a) and 8(b) respectively show the spatial distribution of the turbulent helicity for run 0 and run 1 at each time. For run 1, it is clearly seen that the negative turbulent helicity is dominant at $z > 0$, while the positive turbulent helicity is dominant at $z < 0$. Although the initial condition also has a negative turbulent helicity at $z > 0$ due to the insufficiency of the statistical average, the most part of the turbulent helicity shown in Fig.~\ref{fig:8}(b) is not due to the initial condition but is generated by the system rotation effect at $t>0$. This is suggested by the fact that the turbulent helicity just decays for run 0 with no system rotation [Fig.~\ref{fig:8}(a)]. The same sign of the segregation of the turbulent helicity shown in run 1 was observed in the previous simulations \cite{rd2014,gl1999}, and this result may be obvious from the view point of the inertial wave propagation as stated in Appendix~\ref{sec:b}. In the context of the RANS equation, this spatial distribution of the turbulent helicity antisymmetric about $z=0$ can be also explained by considering the transport equation for the turbulent helicity. It is given by \cite{yy1993}
\begin{align}
\frac{\partial H}{\partial t} = 2\Omega^\mathrm{F}_z \frac{\partial K}{\partial z} + \cdots,
\label{eq:20}
\end{align}
where only the production term is written on the right-hand side for simplicity. Since the turbulent energy is confined near $z=0$ at the initial condition, $\partial K/\partial z$ is negative at $z>0$, while it is positive at $z < 0$. Because $\Omega^\mathrm{F}_z > 0$, a negative $H$ is generated at $z>0$, while a positive $H$ is generated at $z<0$ for rotating cases as observed in Fig.~\ref{fig:8}(b). The spatial distribution of the rotational pressure flux is shown in Fig.~\ref{fig:8}(c). The spatial distribution of the rotational pressure flux is similar to that of the turbulent helicity with negative coefficient at each time. The same tendency is seen for other runs with system rotation. This result suggests that the model expression of the energy flux in terms of the turbulent helicity given by Eq.~(\ref{eq:15}) is qualitatively good. Figure~\ref{fig:9} shows the comparison between the rotational pressure flux and its model given by Eq.~(\ref{eq:15}) with $C_\Omega = 0.03$ for runs 08 and 1 at $2\Omega^\mathrm{F}t=2$ and $4$. The present model predicts the broad spatial distribution of the rotational pressure flux, which cannot be reproduced by the gradient-diffusion approximation. Therefore, the proposed expression is potentially a good candidate for the model of the rotational pressure flux.

\begin{figure}[htp]
\centering
\includegraphics[scale=0.65]{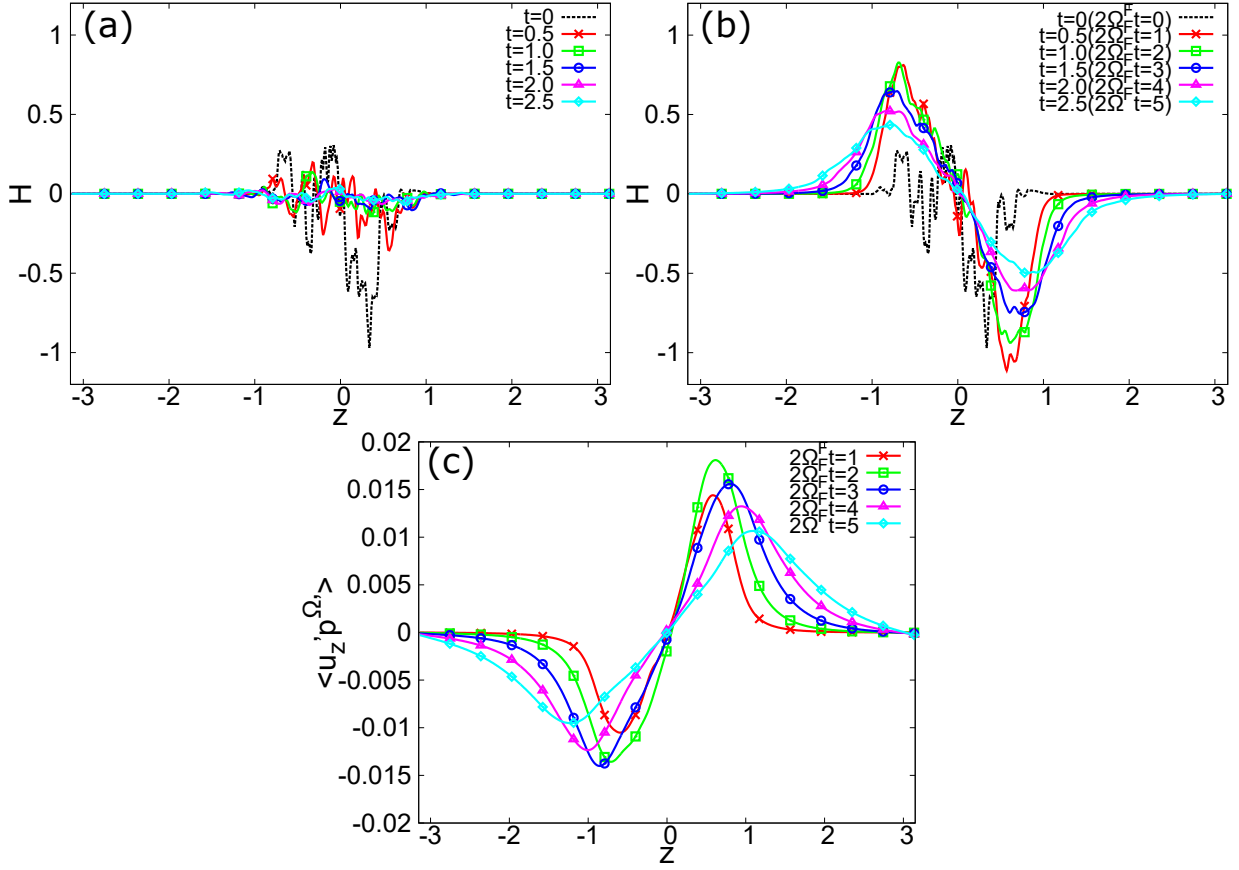}
\caption{Spatial distribution of the turbulent helicity for (a) run 0 and (b) run 1, and (c) the rotational pressure flux for run 1 at each time.}
\label{fig:8}
\end{figure}

\begin{figure}[htp]
\centering
\includegraphics[scale=0.65]{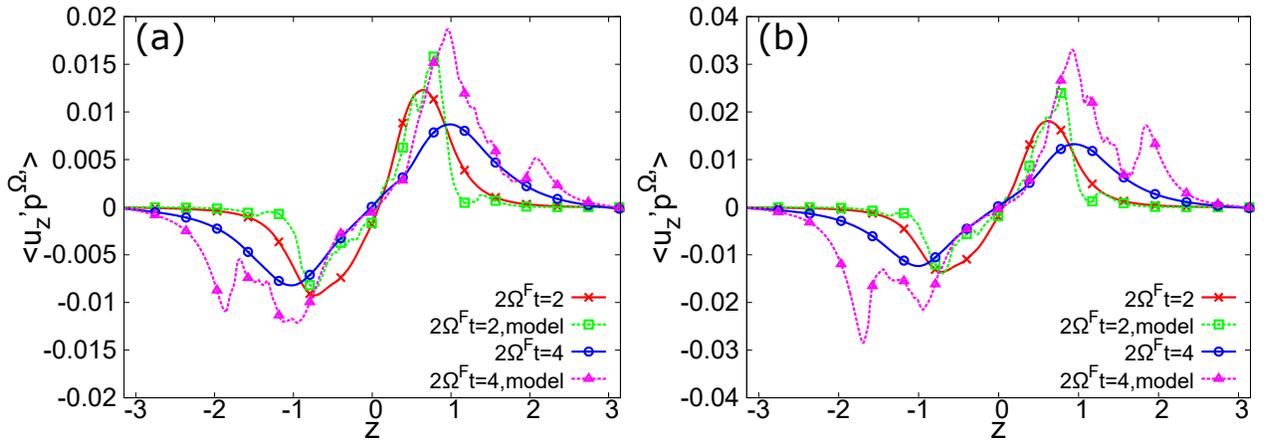}
\caption{Comparison between the rotational energy flux and its model given by Eq.~({\ref{eq:15}}) for (a) run 08 and (b) run 1 at $2\Omega^\mathrm{F} t = 2$ and $4$.}
\label{fig:9}
\end{figure}

\subsection{\label{sec:level4b}Consistency of the model from the analytical view point}

In Fig.~\ref{fig:9}, the prediction by the proposed model at $2\Omega^\mathrm{F} t = 2$ is good, but its accuracy decreases at $2\Omega^\mathrm{F}t = 4$ in the sense that the present model overestimates the DNS value especially at $1 < |z| < 2$. The disagreement at $1 < |z| < 2$ at $2\Omega^\mathrm{F} t = 4$ is partly because the dissipation rate $\varepsilon$ is not adequate to express the coefficient in Eq.~(\ref{eq:15}). The coefficient $K^3/\varepsilon^2$ in Eq.~(\ref{eq:15}) represents the square of the turbulent length scale for the rotational pressure flux. In the RANS modeling, $\varepsilon$ is often interpreted as the energy cascade rate from the large scale to the small scale in addition to the dissipation rate \cite{yoshizawabook}. In the case of fully developed turbulence or a statistically equilibrium state, the energy cascade rate related to the large scales and the dissipation rate related to the small scales are considered to be almost equal; thus the dissipation rate can be used to express the turbulent length scale. However, as seen in Figs.~\ref{fig:5}(b) and \ref{fig:5}(c), the dissipation rate is much less than the rotational pressure diffusion at $1 < |z| < 2$ in this simulation, and the turbulent field is not in an equilibrium state. The dissipation rate is not balanced by the energy cascade rate which is closely related to the turbulent length scale. Therefore, it is possible that the expression $K^3/\varepsilon^2$ overestimates the coefficient of the model given by Eq.~(\ref{eq:15}).

In order to correct the model expression (\ref{eq:15}), we analyze the rotational pressure flux theoretically. Since the flow is homogeneous in $x$ and $y$ directions, the Fourier transformation is applicable in these directions. Here, the Fourier transformation of $q (\mathbf{x})$ in the homogeneous directions is defined as
\begin{subequations}
\begin{align}
q (\mathbf{x}) & = \int \mathrm{d} \mathbf{k}_\perp \hat{q} (\mathbf{k}_\perp,z)
\mathrm{e}^{i \mathbf{k}_\perp \cdot \mathbf{x}_\perp}, 
\label{eq:21a} \\
\hat{q} (\mathbf{k}_\perp, z) & = \frac{1}{(2\pi)^2} \int \mathrm{d} \mathbf{x}_\perp q (\mathbf{x})
\mathrm{e}^{-i \mathbf{k}_\perp \cdot \mathbf{x}_\perp},
\label{eq:21b} 
\end{align}
\end{subequations}
where $\mathbf{k}_\perp = (k_x, k_y)$ and $\mathbf{x}_\perp = (x, y)$. In the case that there is no solid wall, the Poisson equation for the rotational pressure given by Eq.~(\ref{eq:4b}) is solved as
\begin{align}
\hat{p}^\Omega (\mathbf{k}_\perp,z)= 
- 2 \Omega^\mathrm{F}_z \int_{-\infty}^\infty \mathrm{d} z' \frac{1}{2k_\perp} \mathrm{e}^{-k_\perp |z-z'|}
\hat{\omega}_z (\mathbf{k}_\perp, z').
\label{eq:22} 
\end{align}
The rotational pressure flux can be calculated as
\begin{align}
\left< u_z' p^\Omega{}' \right>
& =
\int \mathrm{d} \mathbf{k}_\perp \int \mathrm{d} \mathbf{k}_\perp'
\left< \hat{u}_z' (\mathbf{k}_\perp, z) \hat{p}^\Omega{}' (\mathbf{k}_\perp', z) \right> \mathrm{e}^{i (\mathbf{k}_\perp + \mathbf{k}_\perp')\cdot \mathbf{x}_\perp}
\nonumber \\
& =
\int \mathrm{d} \mathbf{k}_\perp
\Re \left[ \left< \hat{u}_z' (\mathbf{k}_\perp, z) \hat{p}^\Omega{}'{}^* (\mathbf{k}_\perp, z) \right> \right]
\nonumber \\
& = - 2 \Omega^\mathrm{F}_z \int \mathrm{d} \mathbf{k}_\perp \int_{-\infty}^\infty \mathrm{d} z' \frac{1}{2k_\perp} \mathrm{e}^{-k_\perp |z-z'|}
\Re \left[ \left< \hat{u}_z' (\mathbf{k}_\perp,z) \hat{\omega}_z'{}^* (\mathbf{k}_\perp, z') \right> \right],
\label{eq:23} 
\end{align}
where the homogeneity in the $\mathbf{x}_\perp$ direction is used and Eq.~(\ref{eq:22}) is substituted. Equation~(\ref{eq:23}) suggests that the rotational pressure flux can be expressed by using $H_{zz}$($=\langle u_z' \omega_z' \rangle$) with the factor $M^H$ as follows:
\begin{gather}
\left< u_z' p^\Omega{}' \right> = - M^H H_{zz} 2 \Omega^\mathrm{F}_z, 
\label{eq:24} \\
M^H = \frac{1}{H_{zz}} \int \mathrm{d} \mathbf{k}_\perp \int_{-\infty}^\infty \mathrm{d} z' \frac{1}{2k_\perp} \mathrm{e}^{-k_\perp |z-z'|}
\Re \left[ \left< \hat{u}_z' (\mathbf{k}_\perp,z) \hat{\omega}_z'{}^* (\mathbf{k}_\perp, z') \right> \right].
\label{eq:25} 
\end{gather}
Here, $M^H$ has the dimension of the square of the length scale. Modeling $M^H$ requires information on the integral length scales of the turbulent helicity spectrum in the $\mathbf{k}_\perp$ space and the correlation between the velocity and vorticity along the $z$ direction. If the two-point correlation between the velocity and vorticity in the $z$ direction is almost constant within the region at $|z-z'| < 1/k_\perp$, the integral $\int_{-\infty}^\infty \mathrm{d} z' \exp [ - k_\perp |z-z'| ]$ can be calculated separately; $M^H$ can then be written as
\begin{align}
M^H & = \frac{1}{H_{zz}} \int \mathrm{d} \mathbf{k}_\perp k_\perp^{-2}
\Re \left[ \left< \hat{u}_z' (\mathbf{k}_\perp,z) \hat{\omega}_z'{}^* (\mathbf{k}_\perp, z) \right> \right].
\label{eq:26} 
\end{align}
If the integral length scale of the turbulent helicity is comparable to the integral length scale of energy, $M^H$ can be expressed as $M^H \propto (L^K)^2$ where
\begin{align}
L^K & = \frac{1}{2K} \int \mathrm{d} \mathbf{k}_\perp k_\perp^{-1}
\left< \hat{u}_i' (\mathbf{k}_\perp,z) \hat{u}_i'{}^* (\mathbf{k}_\perp,z) \right>.
\label{eq:27} 
\end{align}
Then, the rotational pressure flux can be expressed as
\begin{align}
\left< u_z' p^\Omega{}' \right> = - C_{\Omega L} \left(L^K\right)^2 H 2\Omega^\mathrm{F}_z,
\label{eq:28}
\end{align}
where $C_{\Omega L}$ is a constant. Here, it is also assumed that the turbulent helicity is almost isotropic, $H_{zz} \simeq H/3$. Figure \ref{fig:10} shows the comparison between the rotational pressure flux and the expression (\ref{eq:28}) for runs 08, 1, and 2 at $2\Omega^\mathrm{F} t = 2$ and $4$. Here, $C_{\Omega L} = 0.16$ is adopted. In contrast to the model given by Eq.~(\ref{eq:15}) (Fig.~\ref{fig:9}), the expression (\ref{eq:28}) does not overestimate the exact value at $1 < |z| < 2$ at both time $2\Omega^\mathrm{F} t=2$ and $4$. The same tendency is also shown for run 2 in Fig.~\ref{fig:10}(c). In the case of fully developed turbulence, the energy cascade rate to the small scale is comparable to the dissipation rate, as previously discussed. In such cases, the integral length scale of the energy can be expressed in terms of $K$ and $\varepsilon$ as $L^K \sim K^{3/2}/\varepsilon$. Thus, the expression (\ref{eq:28}) can be rewritten as the model given by Eq.~(\ref{eq:15}). However, the model given by Eq.~(\ref{eq:15}) is not good enough in a non-equilibrium case. In this sense, non-equilibrium effects for the model coefficient $(L^K)^2$ should be incorporated in order to improve the model given by Eq.~(\ref{eq:15}) in future work. Nevertheless, it should be emphasized that the proposed model associated with the turbulent helicity and the system rotation can account for the energy flux enhanced in the direction parallel to the rotation axis, which is not expressed by previous models.

\begin{figure}[htp]
\centering
\includegraphics[scale=0.65]{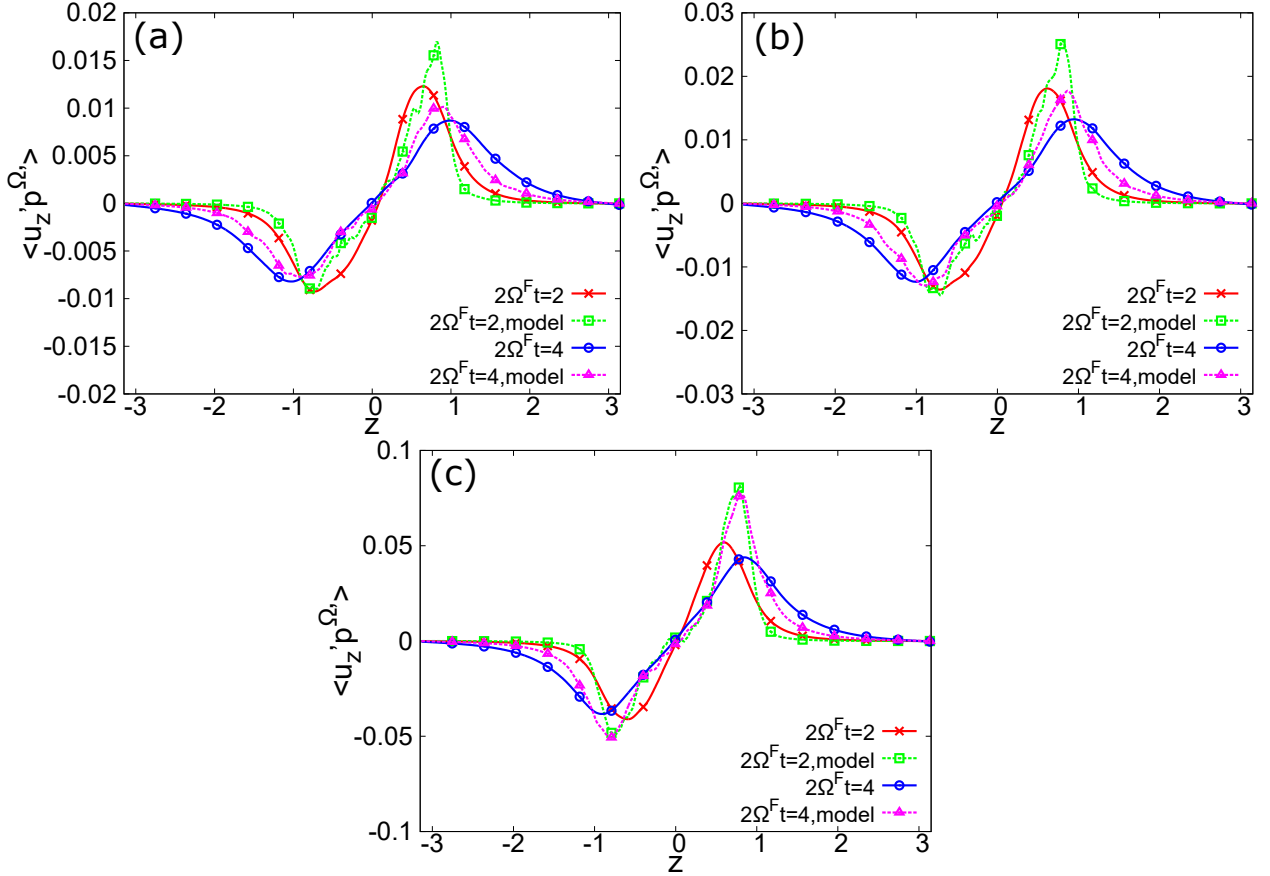}
\caption{Comparison between the rotational pressure flux and the expression (\ref{eq:28}) for (a) run 08, (b) run 1, and (c) run 2 at $2\Omega^\mathrm{F} t = 2$ and $4$.}
\label{fig:10}
\end{figure}

\subsection{\label{sec:level4c}Helical Rossby number}

In this simulation, the energy flux due to the nonlinearity can be predicted using the conventional gradient-diffusion approximation, while the energy flux enhanced by the rotation needs to be predicted by the newly proposed model. However, it is not clear in advance whether the new model is required for simulating general turbulent flows. It would be useful if there existed a criterion for judging the relative importance of the rotational pressure flux in general flows. The conventional Rossby number given by Eq.~(\ref{eq:18}) cannot be used for this purpose because it involves the system rotation, but not the turbulent helicity. Hence, a new non-dimensional parameter which involves both the absolute vorticity and the turbulent helicity is needed as a criterion.

In this study, we define the helical Rossby number $\mathrm{Ro}^H$ as the ratio of the energy flux described by the gradient-diffusion approximation to the energy flux due to the turbulent helicity and the absolute vorticity. By using the expressions (\ref{eq:11}) and (\ref{eq:17}), the helical Rossby number can be defined as
\begin{align}
\mathrm{Ro}^H = 
\left |\frac{(K^2/\varepsilon) \nabla_\parallel K}{(K^3/\varepsilon^2) H \Omega^\mathrm{A}} \right|
= \left |\frac{\varepsilon \nabla_\parallel K}{K H \Omega^\mathrm{A}} \right|,
\label{eq:29} 
\end{align}
where $\nabla_\parallel$ denotes the spatial derivative in the direction of the absolute vorticity and $\Omega^\mathrm{A}$ denotes the absolute value of the absolute vorticity in which the absolute vorticity is already defined in connection with Eq.~(\ref{eq:17}). A major difference between the helical Rossby number and the conventional Rossby number given by Eq.~(\ref{eq:18}) is that the former contains the turbulent helicity. Hence, for non-helical rotating homogeneous turbulence \cite{mnr2001,ymk2011}, the helical Rossby number is infinity although the conventional Rossby number can be small. On the other hand, for inhomogeneous turbulence with the finite turbulent helicity and a mean absolute vorticity, the helical Rossby number has a finite value. In the case of inhomogeneous turbulence accompanied with rotation, the turbulent helicity is often generated \cite{rd2014,gl1999,stepanovetal2018} and one of its generation mechanism is seen in Eq.~(\ref{eq:20}). Figure~\ref{fig:11} shows the distribution of the helical Rossby number given by Eq.~(\ref{eq:29}) for four runs of the rotating cases at $2\Omega^\mathrm{F} t = 2$ in the present simulation. As shown in Fig.~\ref{fig:8}(c), the rotational pressure flux assumes its maximum value near $z = \pm1$. It is clearly seen in Fig.~{\ref{fig:11} that $\mathrm{Ro}^H$ near $z = \pm1$ decreases as the rotation rate increases. In the present simulation, $C_\nu/\sigma_K = 0.22$ is appropriate for the model constant in the gradient-diffusion approximation given by Eq.~(\ref{eq:11}) (see Fig.~\ref{fig:7}), while $C_\Omega = 0.03$ in the model for the rotational pressure flux given by Eq.~(\ref{eq:17}) (see Fig.~\ref{fig:9}). Thus, the ratio of the nonlinear energy flux to the rotational energy flux is estimated as $0.22/0.03 \times \mathrm{Ro}^H \sim 7 \times \mathrm{Ro}^H$. In this sense, $\mathrm{Ro}^H < 1/7$ is a criterion that the energy flux enhanced by the turbulent helicity and the rotation exceeds the energy flux expressed by the gradient-diffusion approximation.

\begin{figure}[htp]
\centering
\includegraphics[scale=0.72]{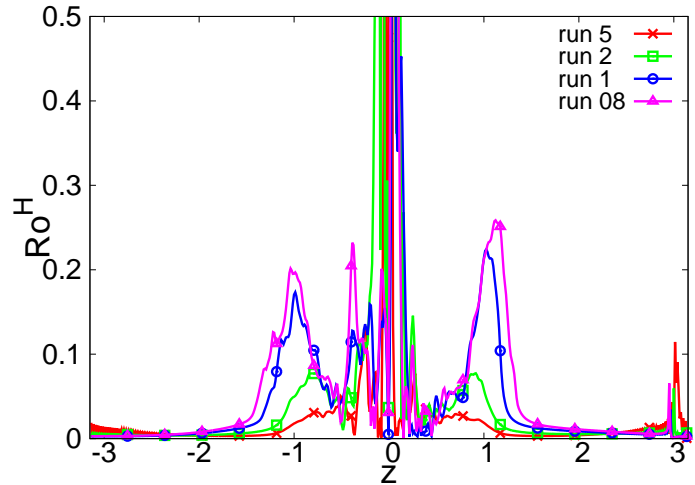}
\caption{Spatial distribution of the helical Rossby number given by Eq.~(\ref{eq:29}) for each run at $2\Omega^\mathrm{F}t = 2$.}
\label{fig:11}
\end{figure}

Although the definition of the helical Rossby number given by Eq.~(\ref{eq:29}) has a physically clear interpretation, it is complex since it contains the spatial derivative in the direction of the rotation axis. Then, we define the simplified helical Rossby number $\mathrm{Ro}^H_\mathrm{s}$ as
\begin{align}
\mathrm{Ro}^H_\mathrm{s} = 
\left |\frac{\varepsilon^2}{K^{3/2} H \Omega^\mathrm{A}} \right|.
\label{eq:30} 
\end{align}
It should be noted that the simplified helical Rossby number still has the same feature as the helical Rossby number given by Eq.~(\ref{eq:29}) in the sense that it has a finite value only when the turbulent helicity is non-zero. Figure~\ref{fig:12} shows the distribution of the simplified helical Rossby number given by Eq.~(\ref{eq:30}) for four runs of the rotating cases at $2\Omega^\mathrm{F} t = 2$ in the present simulation. Although the overall profile is different from Fig.~\ref{fig:11}, it is seen that $\mathrm{Ro}^H_\mathrm{s}$ near $z = \pm1$ decreases as the rotation rate increases. Hence, the simplified helical Rossby number is another candidate criterion for judging the relative importance of the energy flux enhanced by the turbulent helicity and the rotation.

\begin{figure}[htp]
\centering
\includegraphics[scale=0.72]{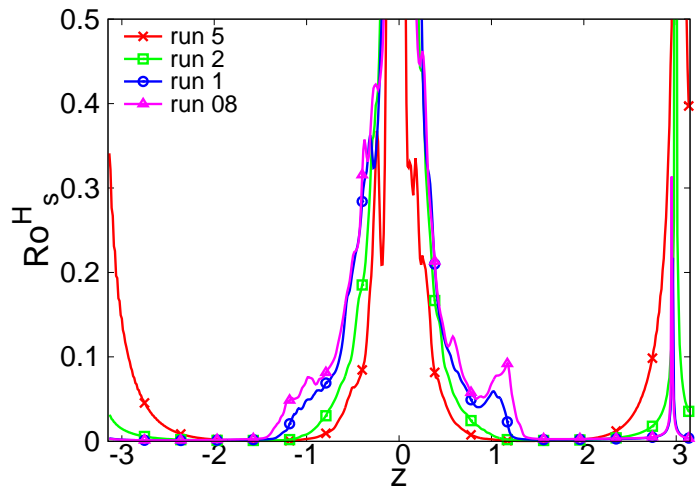}
\caption{Spatial distribution of the simplified helical Rossby number given by (\ref{eq:30}) for each run at $2\Omega^\mathrm{F}t = 2$.}
\label{fig:12}
\end{figure}

\section{\label{sec:level5}Conclusions}

In the case of rotating inhomogeneous turbulence, it is observed that the turbulent energy is rapidly transferred in the direction of the rotation axis in comparison with the non-rotating case \cite{dl1983,dsd2006,kolvinetal2009,rd2014}. The conventional gradient-diffusion approximation of the turbulent energy flux cannot account for this enhancement of the energy flux in the direction parallel to the rotation axis. A new model of the energy flux that represents the energy transport enhanced in the direction parallel to the rotation axis is proposed. The model is associated with the turbulent helicity and the mean absolute vorticity. Its property is similar to the group velocity of inertial waves governed by the linear inviscid equation; the negative instantaneous helicity invokes the energy flux parallel to the rotation axis, while the positive instantaneous helicity invokes the flux anti-parallel to the rotation axis.  In order to assess the validity of the proposed model, a DNS of inhomogeneous turbulence subject to system rotation is performed, whose flow configuration is similar to that proposed by Ranjan and Davidson \cite{rd2014}. It was shown that the rotational pressure diffusion significantly contributes to the diffusion of the turbulent energy. The spatial distribution of the turbulent helicity is similar to that of the rotational pressure flux with negative coefficient, where the rotational pressure flux represents the energy transport enhanced by the system rotation. This result suggests that the new model expressed in terms of the turbulent helicity is qualitatively good. The proposed model agrees well with the exact value at an early stage, while the model overestimates the exact value at a later stage. This overestimation is partly because the turbulent length scale is not adequately expressed by the dissipation rate in a non-equilibrium state. Theoretical analysis revealed that the overestimation can be improved using the integral length scale $L^K$ instead of $K^{3/2}/\varepsilon$ in expressing the model coefficient. In further work, the performance of the model should be assessed in the statistically stationary helical turbulence such as the swirling flow in a straight pipe \cite{kitoh1991,steenbergen} or the swirling jet \cite{stepanovetal2018}. Moreover, we need to improve the model by incorporating non-equilibrium effects through the integral length scale.

In the last section, we introduced the helical Rossby number which represents the ratio of the energy flux described by the gradient-diffusion approximation to that enhanced by the turbulent helicity and the absolute vorticity. The helical Rossby number is different from the conventional Rossby number. The former has a finite value only when the turbulent helicity is non-zero, while the latter can have a finite value even when the turbulent field is non-helical. Turbulent flows associated with the turbulent helicity and the large scale vortex are often encountered in engineering \cite{kitoh1991,steenbergen,stepanovetal2018}, meteorological \cite{lilly1986,nn2010}, and the MHD turbulence \cite{moffattbook,krauseradler,bs2005,hamba2004,wby2016,wby2016-2}. We expect that the helical Rossby number can potentially be utilized as a criterion for judging the relative importance of the energy flux enhanced by the turbulent helicity and the rotation in general turbulent flows.

\begin{acknowledgments}
We wish to acknowledge Dr. Nobumitsu Yokoi for valuable comments and discussion. This work was supported by JSPS KAKENHI Grant Number JP17K06143.
\end{acknowledgments}

\appendix

\makeatletter
\renewcommand{\theequation}{\thesection\arabic{equation}}
\@addtoreset{equation}{section}
\makeatother

\section{\label{sec:a}Analytical modeling of the rotational pressure diffusion}

By using the TSDIA \cite{tsdia}, we obtain a model expression for the rotational pressure diffusion. In the TSDIA, the fast variables $(\bm{\xi}; \tau)$ and the slow variables $(\mathbf{X};T)$ are introduced for space and time variables with a scale parameter $\delta$ as follows:
\begin{align}
\bm{\xi} = \mathbf{x}, \ \ \tau = t, \ \ 
\mathbf{X} = \delta \mathbf{x}, \ \ T = \delta t.
\label{eq:a1}
\end{align}
We assume that the mean values change so slowly that they depend only on the slow variables, $(\mathbf{X};T)$, while the fluctuating fields depend on both the fast and slow variables. This reads as
\begin{align}
q = Q (\mathbf{X};T) + q' (\bm{\xi},\mathbf{X};\tau,T),
\label{eq:a2}
\end{align}
where $q = (u_i,p)$. Under Eq.~(\ref{eq:a1}), the space and time derivatives are expressed as
\begin{align}
\frac{\partial}{\partial x_i} = \frac{\partial}{\partial \xi_i} + \delta \frac{\partial}{\partial X_i}, \ \ 
\frac{\partial}{\partial t} = \frac{\partial}{\partial \tau} + \delta \frac{\partial}{\partial T}.
\label{eq:a3}
\end{align}
The effects of inhomogeneity are included by the derivative expansion in powers of $\delta$. In order to observe the effects of the rotation, the fluctuation fields $q'$ are expanded not only in powers of $\delta$ but also in powers of the rotation parameter $\Omega^\mathrm{F}$ as
\begin{align}
q' (\bm{\xi},\mathbf{X}; \tau, T) = \sum_{n,m=0}^\infty \delta^n |\Omega^\mathrm{F}|^m q^{(nm)} (\bm{\xi},\mathbf{X}; \tau, T).
\label{eq:a4}
\end{align}
The $O(\delta^0 |\Omega^\mathrm{F}|^0)$ field corresponds to the homogeneous non-rotating turbulence. The effects of inhomogeneity and anisotropy are incorporated in the fields of $O(\delta^n)$ with $n \ge 1$, and the effects of the rotation are incorporated in the fields of $O(|\Omega^\mathrm{F}|^m)$ with $m \ge 1$, in a perturbational manner. Here, we assume that the Fourier transformation can be applied to the fast variables $\bm{\xi}$; 
\begin{subequations}
\begin{align}
q (\bm{\xi}, \mathbf{X};\tau, T) & = \int \mathrm{d} \mathbf{k} \ \tilde{q} (\mathbf{k}, \mathbf{X}; \tau, T) 
\mathrm{e}^{i \bm{\xi} \cdot \mathbf{k}},
\label{eq:a5a} \\
\tilde{q} (\mathbf{k}, \mathbf{X};\tau, T) & = \frac{1}{(2\pi)^3} \int \mathrm{d} \bm{\xi} \ q (\bm{\xi}, \mathbf{X}; \tau, T) 
\mathrm{e}^{-i \bm{\xi} \cdot \mathbf{k}}
\label{eq:a5b}.
\end{align}
\end{subequations}
We also assume that the lowest-order field satisfies the following statistical property,
\begin{align}
& \left< \tilde{u}_i^{(00)} (\mathbf{k},\mathbf{X}; \tau, T) \tilde{u}_j^{(00)} (\mathbf{k}',\mathbf{X}; \tau' , T) \right> \nonumber \\
& = \left[ D_{ij} (\mathbf{k}) \frac{E^\mathrm{B} (k,\mathbf{X}; \tau, \tau', T)}{4 \pi k^2} 
- \frac{i}{2} \frac{k_\ell}{k^2} \epsilon_{ij\ell} \frac{E_H^\mathrm{B} (k,\mathbf{X}; \tau, \tau', T)}{4 \pi k^2} \right] \delta (\mathbf{k}+\mathbf{k}') , 
\label{eq:a6}
\end{align}
where $D_{ij} (\mathbf{k}) = \delta_{ij} - k_i k_j / k^2$.  $E^\mathrm{B}$ and $E_H^\mathrm{B}$ denote the spectra of the turbulent energy and helicity of the lowest-order field, respectively. They satisfy the following expression:
\begin{align}
K^\mathrm{B} & = \frac{1}{2} \left< u_i^{(00)} u_i^{(00)} \right> = 
\int_0^\infty \mathrm{d} k \ E^\mathrm{B} (k, \mathbf{X}; \tau, \tau, T) , 
\label{eq:a7} \\
H^\mathrm{B} & = \left< u_i^{(00)} \omega_i^{(00)} \right> = 
\int_0^\infty \mathrm{d} k \ E_H^\mathrm{B} (k, \mathbf{X}; \tau, \tau, T) .
\label{eq:a8}
\end{align}
Up to $O(\delta |\Omega^\mathrm{F}|)$, the pressure diffusion for the turbulent energy transport is calculated as
\begin{align}
\Pi^K & = -\delta \frac{\partial}{\partial X_i} \left[
   \left< u_i^{(00)} p^{(00)} \right> + \left< u_i^{(01)} p^{(00)} \right> 
+ \left< u_i^{(00)} p^{(01)} \right> \right] \nonumber \\
& = \frac{1}{3} \delta \frac{\partial}{\partial X_i} \left[ \int \mathrm{d}k \ k^{-2} E_H^\mathrm{B} (k, \mathbf{X}; \tau, \tau, T) 2\Omega_i^\mathrm{F} \right]
+ O(|u^{(00)}|^3).
\label{eq:a9} 
\end{align}
The expression of the pressure diffusion given by Eq.~(\ref{eq:a9}) was obtained by Inagaki \textit{et al}. \cite{inagakietal2017}. In this study, we obtain the concrete expression of the model in terms of the $K$-$\varepsilon$ model. In the TSDIA, the energy spectrum is expressed by means of the inertial-range form with a low-wavenumber cutoff as \cite{tsdia},
\begin{gather}
E^\mathrm{B} (k) = C_K \varepsilon^{2/3} k^{-5/3},
\label{eq:a10} \\
K^\mathrm{B} = \int_{k^\mathrm{C}}^\infty \mathrm{d} k \ E^\mathrm{B} (k)
= \frac{3}{2} C_K \varepsilon^{2/3} \left(k^\mathrm{C} \right)^{-2/3},
\label{eq:a11} 
\end{gather}
where $C_K$ is the Kolmogorov constant and $k^\mathrm{C}$ denotes the cutoff wavenumber corresponding to the energy-containing scale. Here and hereafter, the dependence of the statistical values on slow variables, $(\mathbf{X};T)$, are omitted for simplicity. The helicity spectrum can be expressed in the following form \cite{yokoi2016};
\begin{align}
E_H^\mathrm{B} (k) = C_H \varepsilon^H \varepsilon^{-1/3} k^{-5/3}.
\label{eq:a12} 
\end{align}
Here, $\varepsilon^H$ denotes the dissipation rate of the turbulent helicity which is defined as
\begin{align}
\varepsilon^H = 2\nu \left< \frac{\partial u_i'}{\partial x_j} \frac{\partial \omega_i'}{\partial x_j} \right>.
\label{eq:a13} 
\end{align}
This helicity spectrum which is proportional to the turbulent helicity dissipation rate was first discussed by Brissaud \textit{et al}. \cite{brissaudetal1973} and $k^{-5/3}$ behavior was numerically observed by the EDQNM approximation \cite{al1977} and DNS \cite{baerenzungetal2008} of homogeneous turbulence. It should be noted that this form of the helicity spectrum (\ref{eq:a12}) is associated with high-Reynolds-number turbulence, but not with weak inertial-wave turbulence \cite{galtier2003}. We choose the helicity spectrum (\ref{eq:a12}) since our focus is on the high-Reynolds-number turbulence. Using the helicity spectrum given by Eq.~(\ref{eq:a12}), we have
\begin{gather}
H^\mathrm{B} = \int_{k^H}^\infty \mathrm{d} k \ E_H^\mathrm{B} (k)
= \frac{3}{2} C_H \varepsilon^H \varepsilon^{-1/3} \left(k^H \right)^{-2/3},
\label{eq:a14} \\
\int_{k^H}^\infty \mathrm{d} k \ k^{-2} E_H^\mathrm{B} (k)
 = \frac{3}{8} C_H \varepsilon^H \varepsilon^{-1/3} \left(k^H \right)^{-8/3}
 = \frac{1}{4} \left(k^H \right)^{-2} H^\mathrm{B},
\label{eq:a15}
\end{gather}
where $k^H$ is the cutoff wavenumber of the lower part of the helicity spectrum. By using Eq.~(\ref{eq:a14}), the wavenumber characterizing the helicity containing scale $k^H$ can be expressed as
\begin{align}
\left( k^H \right)^{-1} = \left( \frac{2}{3C_H} \frac{H^\mathrm{B}}{\varepsilon^H} \right)^{3/2} \varepsilon^{1/2}.
\label{eq:a16} 
\end{align}
If the decaying rate of the turbulent helicity can be estimated by the turbulent time scale $K/\varepsilon$, $\varepsilon^H$ is expressed as \cite{yy1993},
\begin{align}
\varepsilon^H = C_{\varepsilon H} \frac{\varepsilon}{K} H,
\label{eq:a17} 
\end{align}
As the renormalization procedure, we replace $K^\mathrm{B}$ and $H^\mathrm{B}$ by $K$ and $H$, respectively. Using Eqs.~(\ref{eq:a9}) and (\ref{eq:a15})--(\ref{eq:a17}), the model of the pressure diffusion term is expressed as
\begin{align}
\Pi^K = \frac{1}{12} \left( \frac{2}{3C_H C_{\varepsilon H}} \right)^3 \frac{\partial}{\partial x_i} \left[ \frac{K^3}{\varepsilon^2} H 2\Omega_i^\mathrm{F} \right].
\label{eq:a18} 
\end{align}
This expression is the same as Eq.~(\ref{eq:16}).

\section{\label{sec:b}The property of the group velocity of inertial wave}

We briefly explain the property of the group velocity of an inertial wave (see e.g. \cite{davidson} for details). When the system rotation is so rapid that the nonlinear and viscous terms are negligible compared with the Coriolis force term, motion of the fluid is governed by
\begin{align}
\frac{\partial u_i}{\partial t} & =
- \frac{\partial p^\Omega}{\partial x_i}
+ 2 \epsilon_{ij\ell} u_j \Omega^\mathrm{F}_\ell, 
\label{eq:b1}
\end{align}
with the continuity equation (\ref{eq:2}). Note that $p^\mathrm{N} = 0$ under this condition. Taking the curl of each term in Eq.~(\ref{eq:b1}), the vorticity equation in a linear inviscid system is derived as
\begin{align}
\frac{\partial \omega_i}{\partial t} = 2\Omega^\mathrm{F}_j \frac{\partial u_i}{\partial x_j}.
\label{eq:b2}
\end{align}
Taking the curl again and time derivative of Eq.~(\ref{eq:b2}), the wave equation is derived as \cite{davidson}
\begin{align}
\frac{\partial^2}{\partial t^2} \nabla^2 u_i = \left( 2\Omega^\mathrm{F}_j \frac{\partial}{\partial x_j} \right)^2 u_i.
\label{eq:b3}
\end{align}
Substituting the wave solution $u_i = \tilde{u}_i \exp[ i (k_j x_j - \varpi t)]$ to this equation, the frequency $\varpi$ and the group velocity $C^\mathrm{g}_i$ of the inertial wave are obtained as follows:
\begin{align}
\varpi = \pm \frac{2 \Omega^\mathrm{F}_i k_i}{k} , \ \ 
C^\mathrm{g}_i = \frac{\partial \varpi}{\partial k_i} = \pm \frac{2}{k} D_{ij} (\mathbf{k}) \Omega^\mathrm{F}_j,
\label{eq:b4}
\end{align}
where $D_{ij} (\mathbf{k}) = \delta_{ij} - k_i k_j /k^2$. Equation (\ref{eq:b4}) indicates that wave packets in a rotating fluid propagate upward or downward in the direction of the rotation axis. The sign of the frequency and the group velocity of the inertial wave is related to the sign of the instantaneous helicity \cite{moffatt1970}. Substituting the wave solution and the frequency given by Eq.~(\ref{eq:b4}) to the vorticity equation (\ref{eq:b2}), we have
\begin{align}
\tilde{\omega}_i = \mp k \tilde{u}_i,
\label{eq:b5}
\end{align}
where $\tilde{\omega}_i = i \epsilon_{ij\ell} k_j \tilde{u}_\ell$. Thus, the instantaneous helicity is expressed as
\begin{align}
\tilde{u}_i \tilde{\omega}_i^* = \mp k | \tilde{u}_i |^2.
\label{eq:b6}
\end{align}
Hence, the wave packets with negative instantaneous helicity propagate upward, while the packets with positive instantaneous helicity propagate downward, as mentioned in Sec.~\ref{sec:level2c}.

\section{\label{sec:c}Relationship between the inertial waves and the rotational pressure flux}

Here, we consider homogeneous isotropic reflectionally-asymmetric turbulence. In the case of homogeneous turbulence, the diffusion terms such as Eqs.~(\ref{eq:7c})--(\ref{eq:7e}) vanish. However, the energy flux itself can be non-zero. In this situation, the rotational pressure flux is expressed as,
\begin{align}
\left< u_i' p^\Omega{}' \right> = \int \mathrm{d} \mathbf{k} \ 
i \epsilon_{j \ell m} 2\Omega^\mathrm{F}_\ell \frac{k_m}{k^2}
\left<\tilde{u}_i' (\mathbf{k}) \tilde{u}_j'{}^* (\mathbf{k}) \right>.
\label{eq:c1}
\end{align}
In homogeneous isotropic reflectionally-asymmetric turbulence, the velocity correlation is exactly written as
\begin{align}
\left<\tilde{u}_i' (\mathbf{k}) \tilde{u}_j'{}^* (\mathbf{k}) \right>
= D_{ij} (\mathbf{k}) e (\mathbf{k}) - i \epsilon_{ij\ell} \frac{k_\ell}{k} \frac{h (\mathbf{k})}{2k},
\label{eq:c2}
\end{align}
where $e (\mathbf{k})$ and $h (\mathbf{k})$ denote the three-dimensional spectra of the energy and the helicity, respectively. Substituting Eq.~(\ref{eq:c2}) to (\ref{eq:c1}) gives:
\begin{align}
\left< u_i' p^\Omega{}' \right> = - \int \mathrm{d} \mathbf{k} \ 
\frac{2}{k} D_{ij} (\mathbf{k}) \Omega^\mathrm{F}_j \frac{h(\mathbf{k})}{2k}. 
\label{eq:c3}
\end{align}
Note that Eq.~(\ref{eq:23}) for the turbulence which is inhomogeneous in the $z$ direction corresponds to Eq.~(\ref{eq:c3}) for the homogeneous turbulence. If the Rossby number is so small that the governing equation is reduced to the linear inviscid equation (\ref{eq:b1}), the relationship (\ref{eq:b6}) holds. The ensemble average of Eq.~(\ref{eq:b6}) is then written as
\begin{align}
h(\mathbf{k}) = \mp 2k e (\mathbf{k}).
\label{eq:c4}
\end{align}
Substituting Eqs.~(\ref{eq:b4}) and (\ref{eq:c4}) into Eq.~(\ref{eq:c3}), we have
\begin{align}
\left< u_i' p^\Omega{}' \right> = \int \mathrm{d} \mathbf{k} \ 
C^\mathrm{g}_i e (\mathbf{k}).
\label{eq:c5}
\end{align}
This equation can be interpreted as the energy flux due to the group velocity of the inertial waves. Hence, the rotational pressure flux is closely related to the inertial wave propagation. It should be noted, however, the model expressed by Eqs.~(\ref{eq:15}) or (\ref{eq:17}) should be used for fully nonlinear turbulence, while Eq.~(\ref{eq:c5}) is valid only in the linear inviscid regime.

\bibliography{ref}

\end{document}